\documentclass[a4paper,11pt]{article}
\usepackage{pos}
\usepackage{float}

\title{Heavy-light Pseudoscalar Mesons: Light-Front Wave Functions and Generalized Parton Distributions}
\ShortTitle{LFWFs and GPDs of heavy-light pseudoscalar mesons}

\author[a,b]{B. Almeida-Zamora}
\author[c]{J.J. Cobos-Martínez}
\author[d,e]{A. Bashir}
\author[f]{K. Raya}
\author[f]{J. Rodríguez-Quintero}
\author*[b]{J. Segovia}

\affiliation[a]{Departamento de Investigación en Física, Universidad de Sonora,\\
Boulevard Luis Encinas J. y Rosales, 83000, Hermosillo, Sonora, Mexico.}

\affiliation[b]{Departamento de Sistemas Físicos, Químicos y Naturales,\\
Universidad Pablo de Olavide, E-41013 Sevilla, Spain}

\affiliation[c]{Departamento de Física, Universidad de Sonora,\\
Boulevard Luis Encinas J. y Rosales, 83000, Hermosillo, Sonora, Mexico.}

\affiliation[d]{Instituto de Física y Matemáticas, Universidad Michoacana de San Nicolás de Hidalgo,\\ Morelia, Michoacán 58040, México.}

\affiliation[e]{Theory Center, Jefferson Lab,\\ 
Newport News, VA 23606, USA}

\affiliation[f]{Departmento de Ciencias Integradas, Universidad de Huelva,\\
E-21071 Huelva, Spain.}

\emailAdd{bilgai\_almeidaz@hotmail.com}
\emailAdd{jsegovia@upo.es}

\abstract{The internal structure of the lowest-lying pseudo-scalar mesons with heavy-light quark content is thoroughly studied using an algebraic model that has been successfully applied to similar physical observables of pseudoscalar and vector mesons with hidden-flavor quark content, ranging from light to heavy quark sectors. This model is based on constructing simple and evidence-based ansätze for the mesons' Bethe-Salpeter amplitude (BSA) and quark propagator, allowing the Bethe-Salpeter wave function (BSWF) to be computed algebraically. Its projection onto the light front yields the corresponding light-front wave function (LFWF), which provides easy access to the valence-quark Parton Distribution Amplitude (PDA) by integrating over the transverse momentum squared. We leverage our current knowledge of the PDAs of the lowest-lying pseudo-scalar heavy-light mesons to compute their Generalized Parton Distributions (GPDs) via the overlap representation of the LFWFs. From this three-dimensional information, various limits and projections allow us to deduce the related Parton Distribution Functions (PDFs), Electromagnetic Form Factors (EFFs), and Impact Parameter Space GPDs (IPS-GPDs). Whenever possible, we make explicit comparisons with available experimental results and previous theoretical predictions.}

\FullConference{%
QNP2024 - The 10th International Conference on Quarks and Nuclear Physics\\
8-12 July, 2024\\
Facultat de Biologia, Universitat de Barcelona, Barcelona, Spain.
}

\begin{document}
\maketitle

\section{Introduction}
Hadrons are complex bound states of quarks and gluons, governed by Quantum Chromodynamics (QCD). While perturbative QCD is well understood, non-perturbative methods like lattice QCD and Dyson-Schwinger equations (DSEs) are essential for studying hadrons, as quarks and gluons alone cannot explain low-energy formation. Upcoming experiments will provide data to enhance our understanding of hadron structure.

The Dyson-Schwinger approach combines quark propagator DSEs with Bethe-Salpeter equations (BSEs) to compute the Bethe-Salpeter wave function (BSWF) \cite{Roberts:1994dr, Maris:1997tm}. This has led to predictions of meson masses, static properties, and parton distribution amplitudes (PDAs) \cite{Roberts:2011cf}. More complex quantities, such as parton distribution functions (PDFs), generalized parton distributions (GPDs), and transverse momentum distributions (TMDs) are harder to calculate \cite{Shi:2018zqd}.

Simplified algebraic models based on DSEs and BSEs allow for the construction of BSWFs and leading-twist light-front wave functions (LFWFs), effectively predicting properties for various mesons \cite{Albino:2022gzs}. This study uses known PDAs for pseudoscalar mesons \cite{Binosi:2018rht} to compute structural distributions for \(D, D_s, B, B_s,\) and \(B_c\) mesons. This conference proceeding reviews the work presented in Ref.~\cite{Almeida-Zamora:2023bqb} and is organized here as follows: Sec.~\ref{sec:Algebraic Model} outlines the algebraic model framework, the GPD extraction is shown in Sec.~\ref{sec:Internal Structure}, and we conclude with a summary in Sec.~\ref{sec:Summary}.

\section{Algebraic Model}\label{sec:Algebraic Model}
\subsection{Light-Front Wave Function}
The LFWF can be obtained via the light-front projection of the Bethe-Salpeter wave function
\begin{equation}\label{LFWF2}
    \psi_{0^-} (x,p_\perp^2) = \Tr \int \frac{\mathrm{d}^2 k_\parallel}{\pi} \delta (n\cdot k - x n\cdot P) \gamma_5 \gamma \cdot n \chi_{0^-} (p_{-}, P)
\end{equation}
where the Bethe-Salpeter wave function is given by
\begin{align}
  \chi_{0^-} (p_{-}, P) =& S_q(k) \Gamma_{0^-} (p_{-}, P) S_{\bar{Q}}(p-P) \, , \\
    S_{q(\Bar{Q})}(p) =& \left(-i\gamma\cdot p + M_{q(\Bar{Q})}\right) \Delta\left(p^2, M_{q(\Bar{Q})}^2\right) \, , \\
    N_{0^-} \Gamma_{0^-} (p,P) =& i\gamma_5 \int_{-1}^{1} \mathrm{d}w \; \rho_{0^-}(w) \left[ \hat{\Delta}(p^2_{w}, \Lambda_w^2)\right]^{\nu} \label{BSA}
\end{align}
$p_{-} = p - P/2$, $P^{2} = -m^{2}_{0^-}$ is the mass squared of the meson, $S_{q(\Bar{Q})}$ is the light-quark (heavy-antiquark) propagator and $\Gamma_{0^-}$ the Bethe-Salpeter amplitude; $p_w = p + \frac{w}{2}P $, $\Delta(s,t)=(s+t)^{-1}$ and $\hat{\Delta}(s,t)=t\Delta(s,t)$ and $k_w = k + \frac{w}{2} P$. The notation $J^{P} = 0^{-}$ indicates the spin-parity quantum numbers, $ M_{q(\Bar{Q})}$ is the dynamically-dressed quark (antiquark) mass, $N_{0^-}$ is a normalisation constant, $\rho(w)$ the spectral density function and $\Lambda_w^2= \Lambda^2(w) = M_q^2 - \frac{1}{4} (1 - w^2) m_{0^-}^2 + \frac{1}{2} (1-w) (M_{\bar{Q}}^2 - M_q^2)$. The trace in Eq. \eqref{LFWF2} is taken over color and spinor indices, $n$ is a light-like vector, such that $n^2 = 0 $ and $n \cdot P = - m_{0^-}$. The variable $x$ corresponds to the light-front momentum fraction carried by the quark. From the definition of Mellin's moments, $\expval{x^m} = \int_0^1 dx \; x^m \, \psi (x)$, and some techniques of Feynman parametrisation, the uniqueness of the Mellin moments allows us to identify the LFWF as
\begin{align}
N_{0^-}\psi_{0^-}(x,p_\perp^2) =& 12 \left[ \int_{-1}^{1-2x}dw \left( \frac{x}{1- w}\right)^\nu + \int_{1-2x}^1dw \left(\frac{\Bar{x}}{1-w}\right)^\nu \right]  \frac{2^\nu (x M_{\bar Q} + \Bar{x}M_q) \tilde{\rho}_{0^-}^\nu (w)}{\left[p_\perp^2 + \Lambda_{1-2x}^2\right]^{\nu+1}} \,.  
\end{align}
where $\Bar{x}=(1-x)$ and $\tilde{\rho}_{0^-}^\nu(w) = \Lambda_w^{2\nu} \, \rho_{0^-}(w)$. With the LFWF at hand, $\psi_{0^-} (x,p_\perp^2)$, its integration over $p^2_\perp$-dependence yields the PDA
\begin{equation}
    f_{0^-} \phi_{0^-} (x) = \int \frac{d^2 p_\perp}{16\pi^3} \psi_{0^-} (x,p_\perp^2)
\end{equation}\label{eq:LFWFfinal}
where $f_{0^-}$ is the leptonic decay constant. We arrive at the following algebraic relation
\begin{equation}\label{LFWF}
\psi_{0^-} (x,p_\perp^2) = 16\pi^2 f_{0^-} \frac{\nu (\Lambda_{1-2x}^2)^{\nu}}{(p_\perp^2 + \Lambda_{1-2\alpha}^2)^{\nu+1}}  \phi_{0^-} (x) 
\end{equation}
This result is a merit of the algebraic model. Note also that, throughout this manuscript, we shall employ dimensionless and unit normalized PDAs, \textit{i.e.} $\int_0^1 dx \phi_{0^-} (x) =1  $.
The tools for calculating LFWFs from PDAs are now available. The PDAs used here are based on the Continuum Schwinger Function Method (CSM) and computed using Rainbow Ladder (RL) truncation within a reliable domain and extrapolated via the Schlessinger point method (SPM). The distribution amplitudes for $D, D_s, B, B_s$, and $B_c$ mesons are provided by Ref. \cite{Binosi:2018rht}. Using these PDAs, we will calculate the LFWFs for the $D$- and $B$-mesons, as linked to the corresponding PDAs in Eq. \eqref{eq:LFWFfinal} (details in Ref. \cite{Almeida-Zamora:2023bqb}).

\begin{figure}[!t]
    \centering
    
    \begin{subfigure}[t]{0.4\textwidth}
        \centering
        \includegraphics[width=\textwidth]{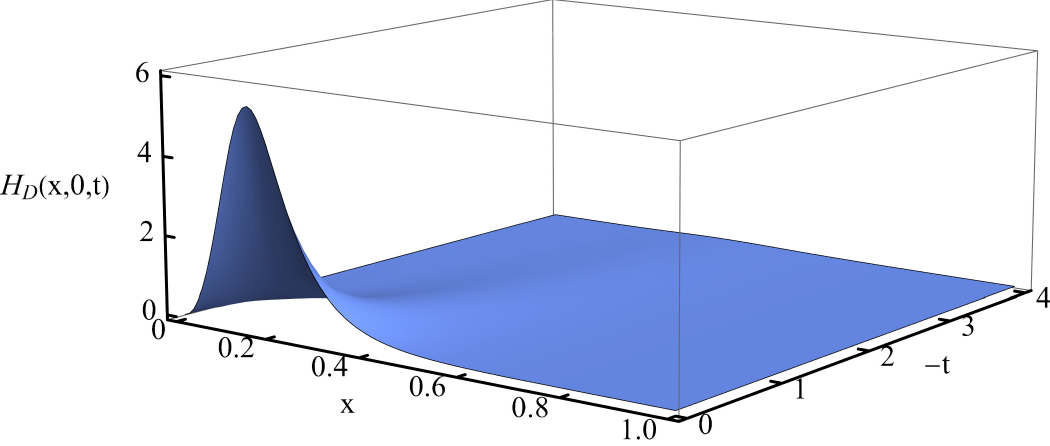}
        \caption*{$D$-meson GPD.}
        \label{fig:GPD_Dn}
    \end{subfigure}
    \hfill
    \begin{subfigure}[t]{0.4\textwidth}
        \centering
        \includegraphics[width=\textwidth]{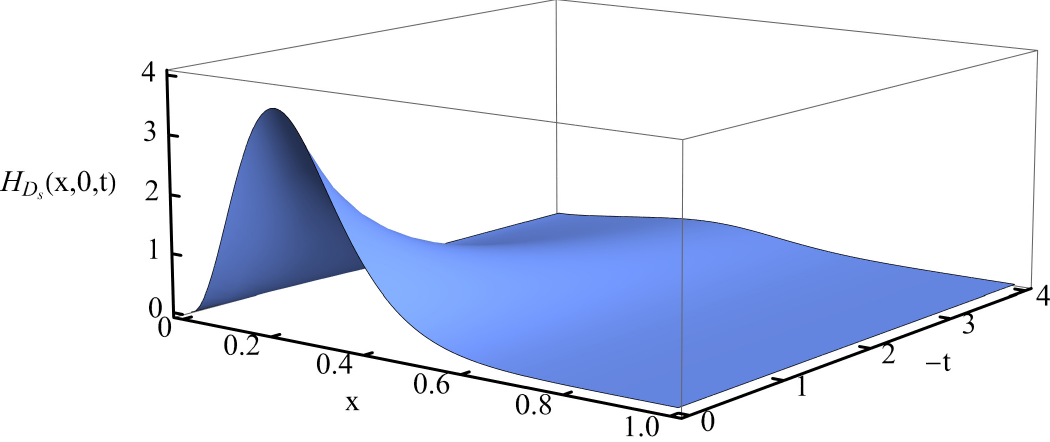}
        \caption*{$D_s$-meson GPD.}
        \label{fig:GPD_Ds}
    \end{subfigure}
    
    \vspace{0.50cm}
    
    \begin{subfigure}[t]{0.32\textwidth}
        \centering
        \includegraphics[width=\textwidth]{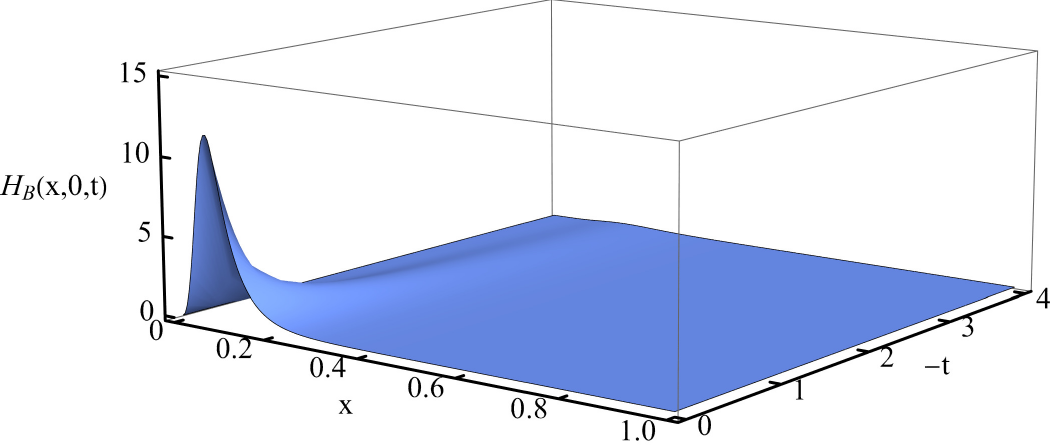}
        \caption*{$B$-meson GPD.}
        \label{fig:GPD_Bn}
    \end{subfigure}
    \hfill
    \begin{subfigure}[t]{0.32\textwidth}
        \centering
        \includegraphics[width=\textwidth]{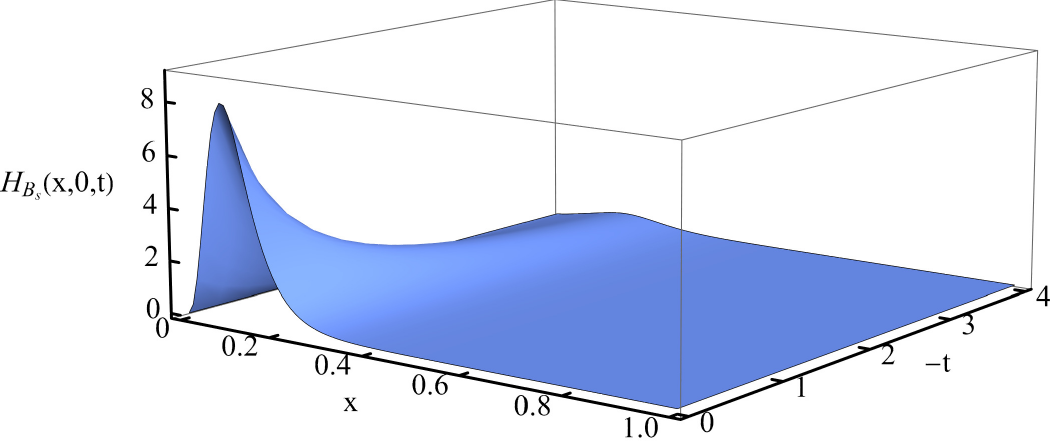}
        \caption*{$B_s$-meson GPD.}
        \label{fig:GPD_Bs}
    \end{subfigure}
    \hfill
    \begin{subfigure}[t]{0.32\textwidth}
        \centering
        \includegraphics[width=\textwidth]{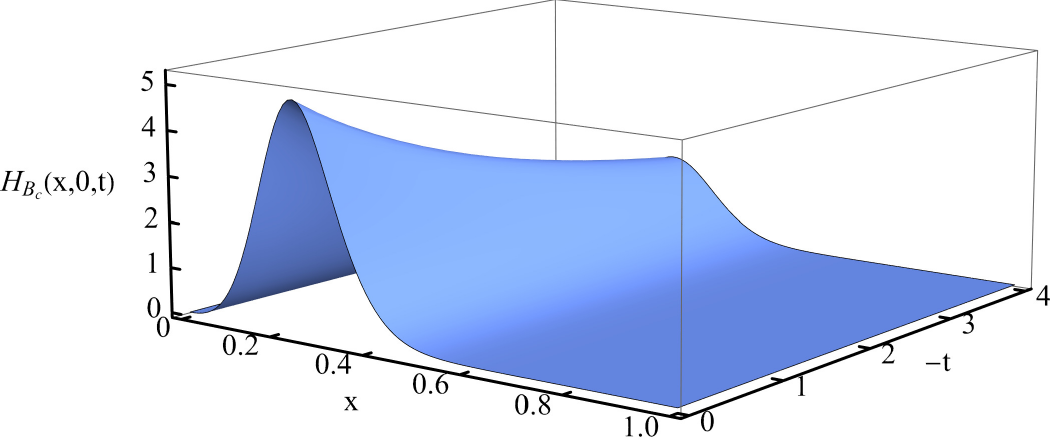}
        \caption*{$B_c$-meson GPD.}
        \label{fig:GPD_Bc}
    \end{subfigure}
    \caption{ Valence quark GPDs obtained from Eq. \eqref{eq:GPDfinal} for $\xi=0$. \emph{Upper panel:} $D$-meson GPD and \emph{lower panel:} $B$-meson GPDs. Mass units in GeV.}
        \label{fig:GPDsmesons}
\end{figure}

\section{Internal Structure of Pseudoscalar Heavy-light Mesons}\label{sec:Internal Structure}
\subsection{Generalised Parton Distributions}
Calculating GPDs directly from QCD is difficult, but their properties inform modeling strategies. Two common methods for evaluating GPDs are the double distribution \cite{Radyushkin:1997ki} and overlap \cite{Diehl:2003ny} representations. The double distribution method maintains polynomiality but complicates the management of positivityy, while the overlap representation preserves positivity but complicates polynomiality management. The covariant extension approach \cite{Mezrag:2014jka, Chouika:2017rzs, Chouika:2017dhe, Chavez:2021llq, Chavez:2021koz} seeks to create GPDs that satisfy both properties by constructing positive DGLAP GPDs using the overlap representation of LFWFs. This makes the algebraic model effective for calculating GPDs for pseudoscalar heavy-light mesons. The valence quark GPD can be obtained as
\begin{equation}\label{GPDdef}
    H_{0^-}(x,\xi,t) = \int \frac{d^2 p}{16\pi^3} \psi_{0^-}^{*}(x^-, (\boldsymbol{p_\perp^-})^2) \psi_{0^-}(x^+, (\boldsymbol{p_\perp^+})^2),
\end{equation}
where $x^\pm = \frac{x \pm \xi}{1 \pm \xi}$ and $\boldsymbol{p_\perp^\pm}= \boldsymbol{p_\perp} \mp \frac{\Delta_\perp}{2} \frac{1-x}{1 \pm \xi}$. Here, $t = - \Delta^2 = - (p - p')^2$ is the momentum transfer, with $p$ ($p'$) being the initial (final) meson moment. Additionally, the skewness is defined as $\xi= -\frac{n\cdot \Delta}{2 n \dot P}$, where $P=\frac{p+p'}{2}$ is the total momentum of the system. $x$ and $\xi$ have support on $\left[-1, 1\right]$, but the overlap representation is only considered in the DGLAP region, $\lvert x \rvert >  \lvert \xi \rvert$. Substituting the LFWF in Eq. \eqref{LFWF} into Eq. \eqref{GPDdef} we can obtain the final expression for the valence quark GPD is given by
\begin{align}\label{eq:GPDfinal}
    \mathcal{H}_{0^-} (x,\xi,t)=& \frac{\Gamma(2\nu +2)}{\Gamma^2 (\nu + 1)}  (f_{0^-} \nu)^2\Lambda_{1-2x^-}^{2\nu} \Lambda_{1-2x^+}^{2\nu} \phi_{0^-}(x^-) \phi_{0^-}(x^+)\int_0^1 du \frac{u^\nu (1-u)^u}{\left[ \mathbf{M}^2(u) \right]^{2\nu +1}}
\end{align}
where $\mathbf{M}^2(u) = c_2 u^2 + c_1 u + c_0 $, with $c_2 = \frac{(1-x)^2}{(1-\xi^2)^2} t$, $c_1 = - \frac{(1-x)^2}{(1-\xi^2)^2} t + \Lambda_{1-2x^+}^{2\nu} - \Lambda_{1-2x^-}^{2\nu}$ and $c_0 = \Lambda_{1-2x^-}^{2\nu} $. Figure~\ref{fig:GPDsmesons} shows the GPDs for $D$- and $B$-mesons at $\xi=0$. The $D$-meson GPD (upper panel) decreases sharply with momentum transfer, nearly vanishing beyond $-t \approx 1\,\text{GeV}^2$, with its $x$-dependence concentrated at $x \lesssim 0.5$ and peaking around $x = 0.2$. The $D_s$-meson GPD behaves similarly but more smoothly, approaching zero above $2\,\text{GeV}^2$. In contrast, the $B$, $B_s$, and $B_c$ mesons exhibit a strong momentum transfer dependence, with narrower $x$-ranges and peaks shifting toward larger longitudinal momentum fractions. This trend is particularly evident in the $B$ and $B_s$ mesons, reflecting the influence of quark-antiquark content and mass differences. The $B_c$-meson, with both heavy quarks, shows the smoothest $t$-dependence, remaining significant for $-t \gtrsim 4\,\text{GeV}^2$ and peaking around $x \approx 0.3$.

\subsection{Electromagnetic Form Factors}
The elastic electromagnetic form factor (EFF) is obtained from the zeroth moment of the GPD
\begin{equation}
F_{0^-}^q(t) = \int_{-1}^1 dx\; H_{0^-}^q(x,\xi,t) \,,
\label{eq:EFFq}
\end{equation}
an analogous expression holds for the antiquark $\bar{Q}$. The complete meson's EFF is given by
\begin{equation}
F_{0^-}(t) = e_q \, F_{0^-}^q(t) + e_{\bar{Q}} \, F_{0^-}^{\bar{Q}}(t) \,, 
\label{eq:MFF1}
\end{equation}
where $e_{q(\bar{Q})}$ is the electric charge of the light quark (heavy antiquark) in units of the positron charge. Due to the polynomiality property of the GPD, the EFF does not depend on $\xi$, therefore, one can simply take $\xi=0$. A Taylor expansion around $t\approx 0$ yields a relation between the EFF and the charge radius:
\begin{eqnarray}
\label{eq:chargeradius}
(r_{0^-}^q)^2 &=&\left. -6 \frac{d F_{0^-}^q(t)}{dt} \right|_{t=0}\;,
\end{eqnarray}
where $r_{0^-}^q$ denotes the contribution of the quark $q$ to the meson charge radius, $r_{0^-}$. On the other hand, from Eq. \eqref{eq:GPDfinal}, by taking $\xi=0$ and expanding $\mathbb{M}^2(u)$ around $-t\approx 0$, an algebraic solution for Eq.~\eqref{eq:GPDfinal}  is obtained (see details in Ref. \cite{Almeida-Zamora:2023bqb}). When integrated with respect to $x$ this leads to a semi-analytical expression for $r_{0^-}^q$:
\begin{align}
(r_{0^-}^q)^2=&6 \, \int_0^1 dx\;\hat{f}_{0^-}^q(x)q_{0^-}(x) \,, \label{eq:crquark} \\
\hat{f}_{0^-}^q(x) =&\frac{c_\nu^{(1)}(1-x)^2}{\Lambda_{1-2x}^2}z \label{eq:fqEFF}
\end{align}
and $q_{0^-}(x) \equiv H_{0^-}(x,0,0)$  defines the valence quark parton distribution function (PDF). Eq. \eqref{eq:crquark} shows that the charge radius is tightly connected with the hadronic scale PDF, and thus with the corresponding PDA. The antiquark result is obtained analogously, with the shift $x \rightarrow (1-x)$. Summing up the quark and antiquark contributions, the meson charge radius reads:
\begin{equation}
\label{eq:crgeneral}
r_{0^-}^2= e_q (r_{0^-}^q)^2 + e_{\bar{Q}} (r_{0^-}^{\bar{Q}})^2\,.
\end{equation}
\begin{figure}[!t]
\centering
\includegraphics[width=0.45\textwidth]{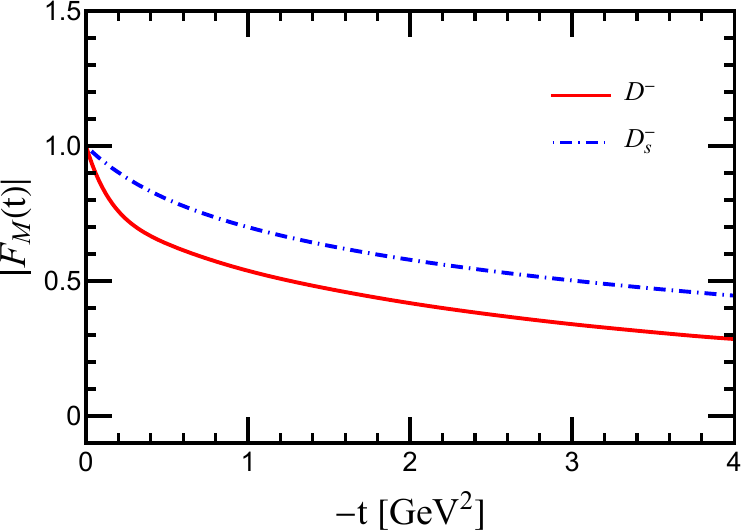}
\hspace*{0.50cm}
\includegraphics[width=0.45\textwidth]{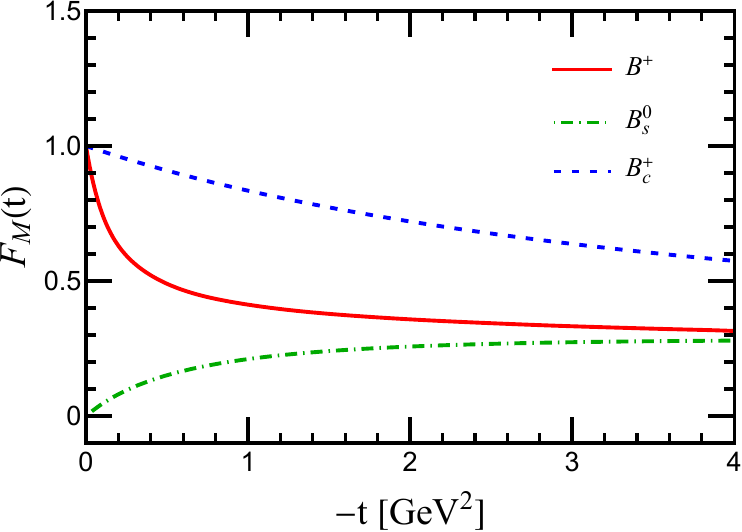}
\caption{\label{fig:EFFs} \emph{Left panel:} EFFs of pseudoscalar charmed mesons. \emph{Right panel:} Analogous, bottom mesons. The EFFs correspond to charged states: $D^-=d\bar c$, $D_s^-=s\bar c$, $B^{+}=u\bar b$, $B_s=s\bar b$ and $B_c^+=c\bar b$.}
\end{figure}
\begin{table}[!t]
\centering
\caption{\label{tab:EMradii} Charge radii (in fm) of the lowest-lying pseudo-scalar heavy-light mesons.}
\begin{tabular}{lr|r|r|r|r|r}
& & $D^{-}$ & $D_{s}^{-}$ & $B^{+}$ & $B_s^{0}$ & $B_c^{+}$ \\
\hline
$|r_{0^{-}}|$ & & 0.680 & 0.372 & 0.926 & 0.345 & 0.217 \\
Covariant CQM & \cite{Moita:2021xcd} & 0.505 & 0.377 & - & - & - \\
PM & \cite{Das:2016rio} & -  & 0.460 & 0.730 & 0.460 & - \\
LFQM & \cite{Hwang:2001th} & 0.429 & 0.352 & 0.615 & 0.345 & 0.208 \\
CI & \cite{Hernandez-Pinto:2023yin} & - & 0.260 & 0.340 & 0.240 & 0.170 \\
Lattice & \cite{Li:2017eic} & 0.450(24) & 0.465(57) & - & - & - \\
Lattice & \cite{Can:2012tx} & 0.390(33) & - & - & - & - \\
\end{tabular}
\end{table}
Figure~\ref{fig:EFFs} shows the EFFs for charmed (left panel) and bottom mesons (right panel). Generally, EFF decreases more smoothly with higher momentum transfer when valence quark mass differences are smaller. Mesons in the same heavy quark sector exhibit similar asymptotic decreases. This trend aligns with findings from Refs.~\cite{Moita:2021xcd, Das:2016rio, Hwang:2001th, Hernandez-Pinto:2023yin}, though direct comparisons are complicated by different hadronic scales. Table~\ref{tab:EMradii} presents charge radius data from lattice QCD and our results, which are consistent for the $D_s$, $B_s$, and $B_c$ mesons, but show larger charge radii for the $D$ and $B$ mesons, likely due to the significant mass difference between their valence quarks.

\subsection{Impact parameter space GPD}
\begin{figure}[!t]
    \centering
    
    \begin{subfigure}[t]{0.32\textwidth}
        \centering
        \includegraphics[width=\textwidth]{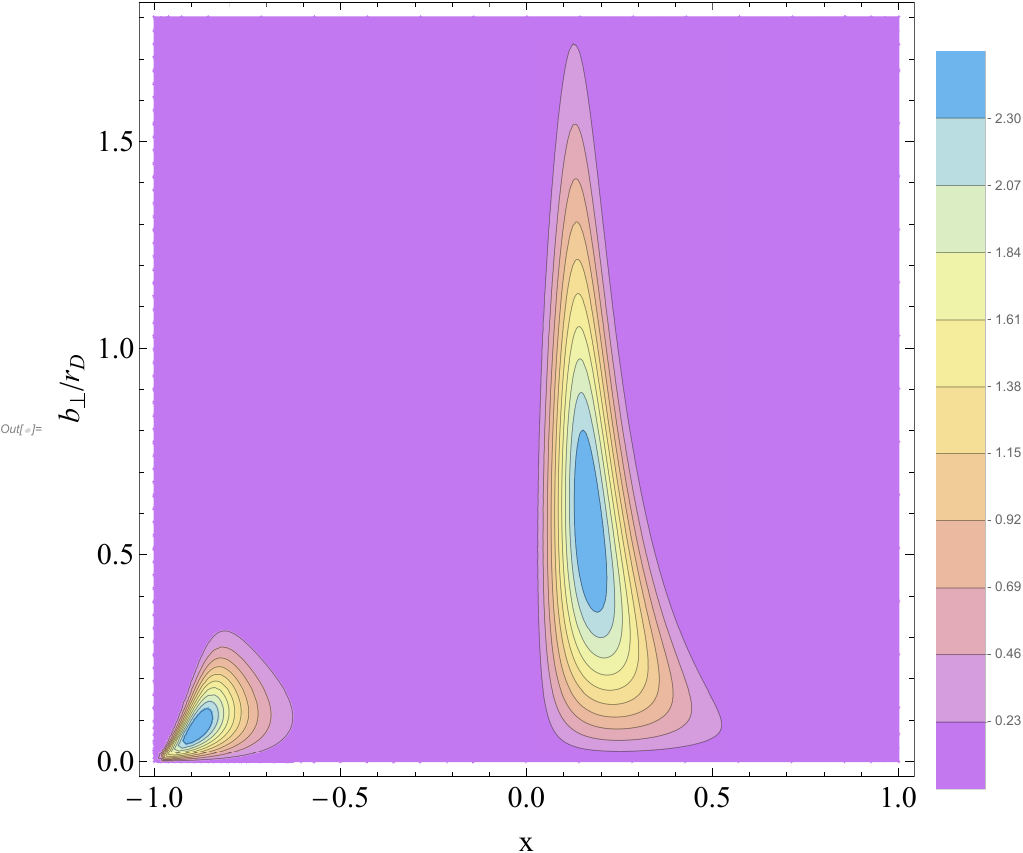}
        \caption*{$D$-meson IPS-GPD.}
    \end{subfigure}
    \hfill
    \begin{subfigure}[t]{0.32\textwidth}
        \centering
        \includegraphics[width=\textwidth]{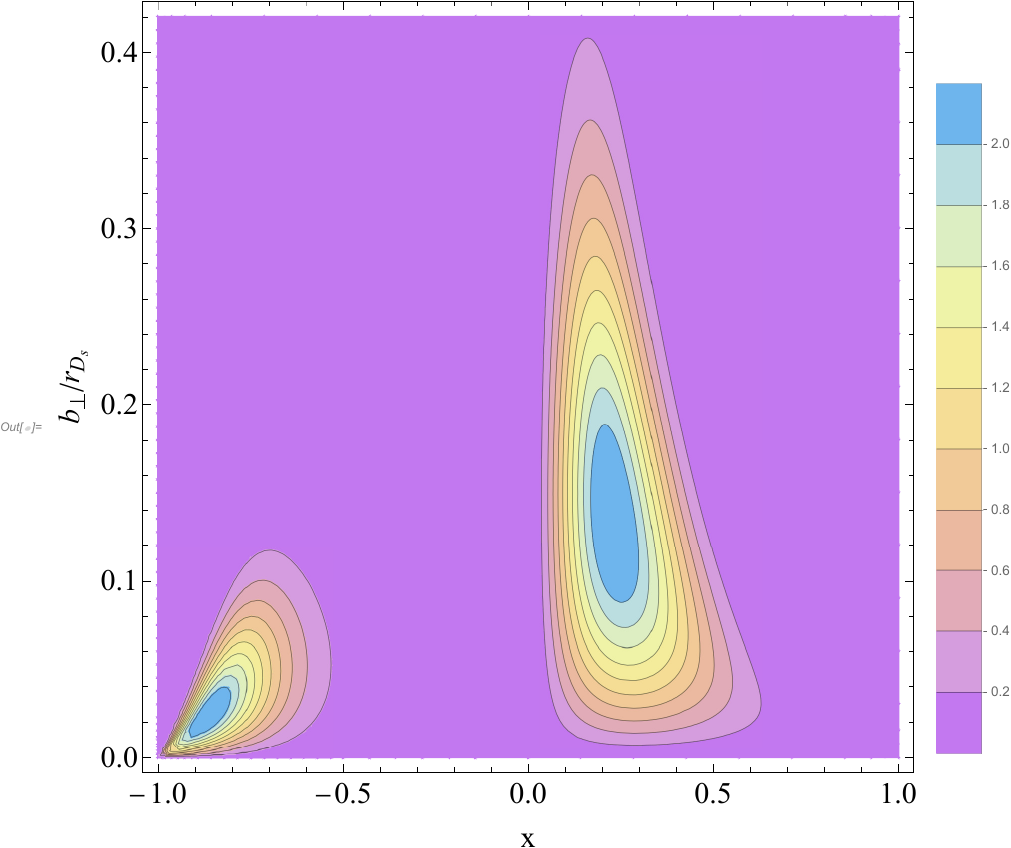}
        \caption*{$D_s$-meson IPS-GPD.}
    \end{subfigure}
    
    \vspace{0.50cm}
    
    \begin{subfigure}[t]{0.32\textwidth}
        \centering
        \includegraphics[width=\textwidth]{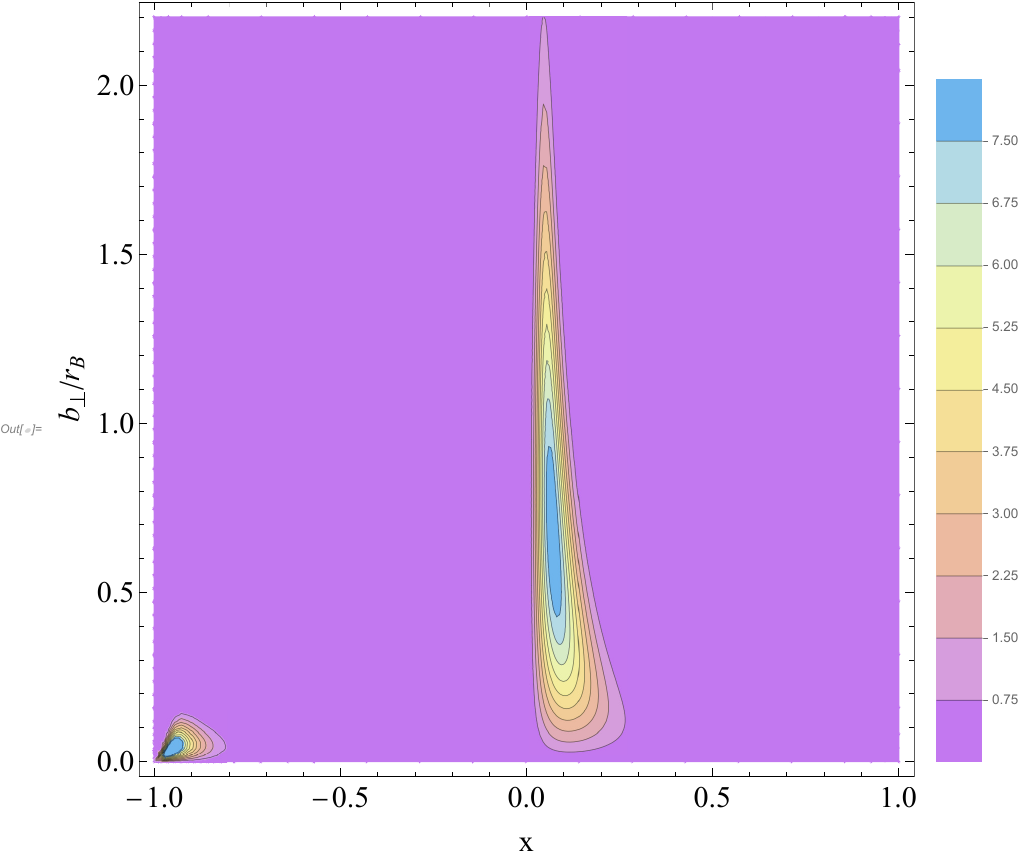}
        \caption*{$B$-meson IPS-GPD.}
    \end{subfigure}
    \hfill
    \begin{subfigure}[t]{0.32\textwidth}
        \centering
        \includegraphics[width=\textwidth]{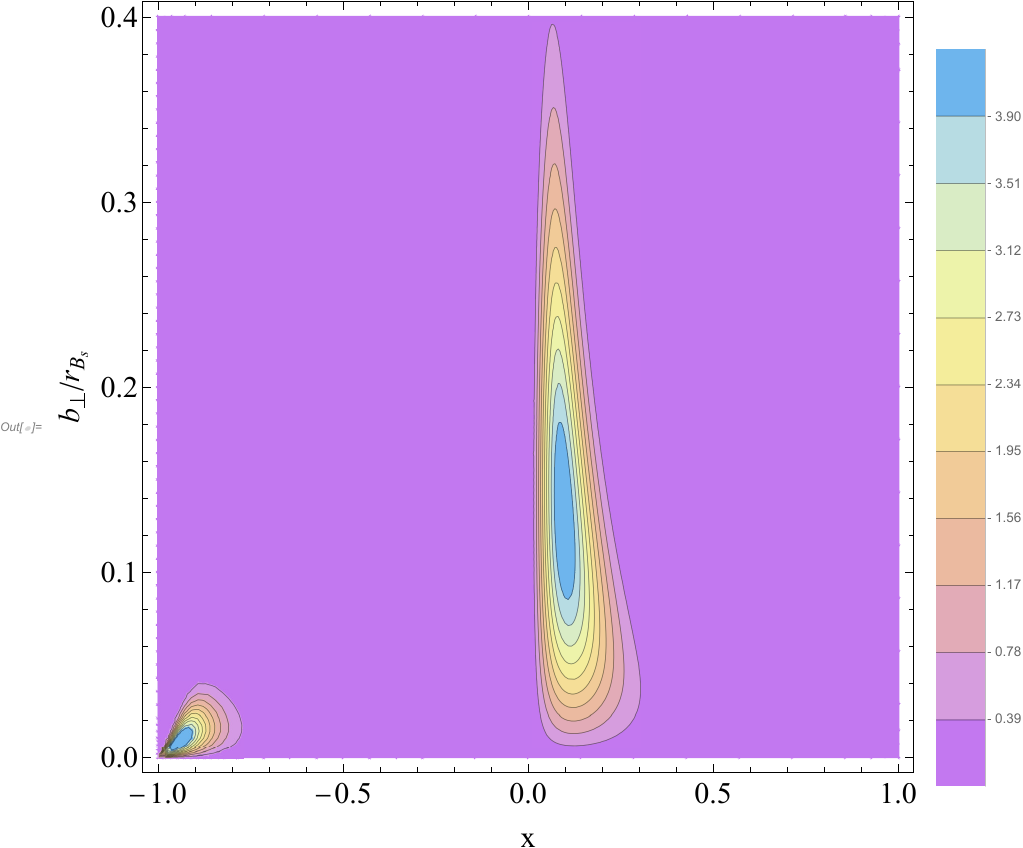}
        \caption*{$B_s$-meson IPS-GPD.}
    \end{subfigure}
    \hfill
    \begin{subfigure}[t]{0.32\textwidth}
        \centering
        \includegraphics[width=\textwidth]{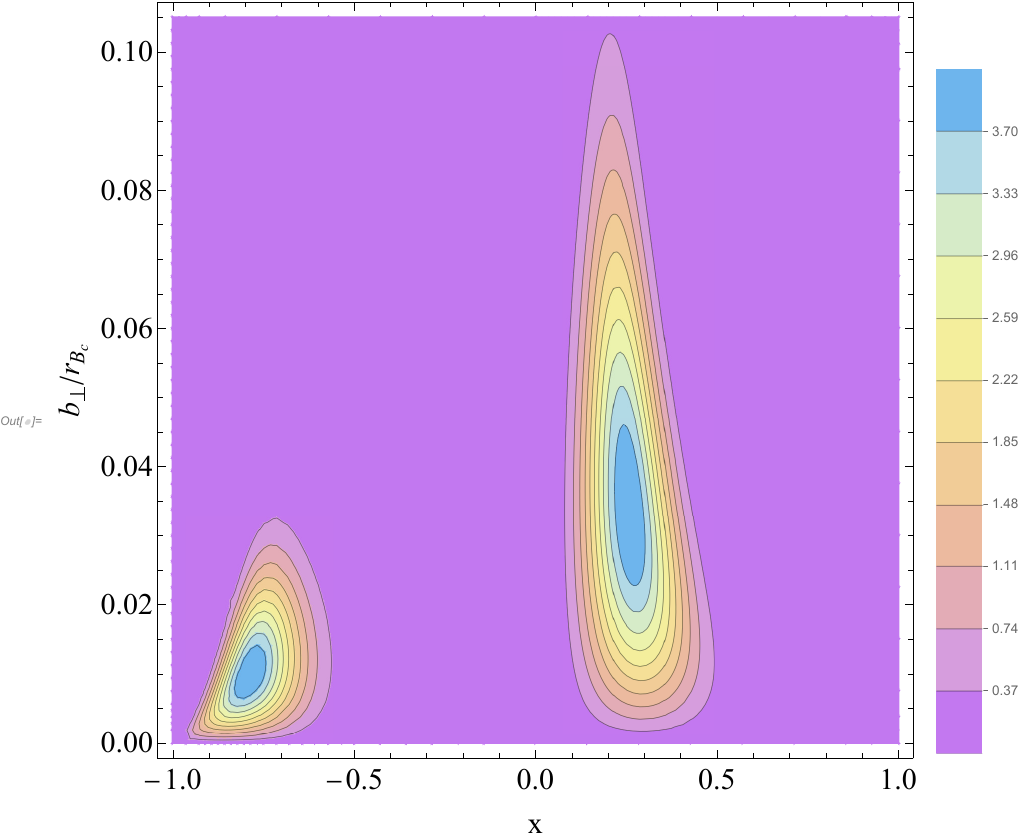}
        \caption*{$B_c$-meson IPS-GPD.}
    \end{subfigure}
    \caption{\label{fig:IPS-GPD_Dmesons} For illustrative purposes, we have considered the convenient representation of Eq.~\eqref{eq:bqblack}. \emph{Upper panel:} $D$-mesons results. \emph{Lower panel:} analogous results for the $B$-mesons.}
\end{figure}

The IPS-GPD distribution represents the probability density of locating a parton with momentum fraction $x$ at a transverse distance $b_{\perp}$ from the meson's center of transverse momentum. It can be directly derived by performing the Fourier transform of the zero-skewness GPD:
\begin{eqnarray}
u_{0^-}(x,b_{\perp}^2)= \int_0^{\infty}\frac{d\Delta}{2\pi }\Delta J_0 (b_{\perp} \Delta) H_{0^-}^q(x,0,t) \,,
\end{eqnarray} 
where $J_0(z)$ is the zeroth Bessel function of the first kind. From the representation of GPD within light-front holographic QCD approach \cite{Chang:2020kjj,deTeramond:2018ecg}, we obtained an analytic expression for the IPS-GPD
\begin{eqnarray}
\label{eq:IPSux}
u_{0^-}^q(x,b_{\perp}^2)= \frac{q_{0^-}(x)}{4 \pi \hat{f}_{0^-}^{q}(x)} \text{exp}\left[-\frac{b_\perp^2}{4\hat{f}_{0^-}^{q}(x)}\right]\;.
\end{eqnarray}
This contains an explicit dependence on the PDF and reveals a clear interrelation between a parton's momentum and its spatial distribution within a meson. Considering the mean-squared transverse extent (MSTE):
\begin{eqnarray}
\expval{b_\perp^2(x) }_{0^-}^q &=& \frac{1}{r_{0^-}}\int_0^\infty db_\perp\, b_\perp^2\, \textbf{b}_{0^-}^q(x,b_\perp) \,, \\
\textbf{b}_{0^-}^q(x,b_\perp) &:=& 2\pi r_{0^-} b_\perp u_{0^-}^q(x,b_\perp)  \label{eq:bqblack}
\end{eqnarray}
the IPS-GPD, defined in Eq.~\eqref{eq:IPSux}, yields the plain relation:
\begin{equation}
\expval{ b_\perp^2}_{0^-}^q = 4\int_0^1 dx\,\hat{f}_{0^-}^{q}(x)q_{0^-}(x) \,.
\end{equation}
Integrating over $x$ and comparing with Eq.~\eqref{eq:crgeneral}, one can obtain the expectation value:
\begin{equation}
\expval{b_\perp^2}_{0^-}^q =  \frac{2}{3}r_{0^-}^2  \left[ \frac{(r_{0^-}^q)^2}{e_q(r_{0^-}^q)^2+e_{\bar{Q}}(r_{0^-}^{\bar{Q}})^2} \right] \,. \label{eq:aveMSTE}
\end{equation}
The expectation value of the MSTE is directly related to the meson charge radius. Figure~\ref{fig:IPS-GPD_Dmesons} shows the IPS-GPD for \(D\)-mesons (upper panel). The quark is in the \(x>0\) domain and the antiquark in \(x<0\). The heavy antiquark remains near the center of transverse momentum, while the light quark is typically found at distances of \(0.6 \times r_D\) and \(0.14 \times r_{D_s}\) for the \(D\) and \(D_s\) mesons, respectively. As the constituent quark mass increases, the quark increasingly dictates the center of transverse momentum, leading to wider distributions in \(x\), narrower distributions in \(b_\perp\), and smaller maxima. Similar trends for charmed mesons apply to bottom mesons, where the heavy antiquark also significantly influences the center of transverse momentum. The light quark is located at \(0.65 \times r_B\), \(0.13 \times r_{B_s}\), and \(0.035 \times r_{B_c}\) for the \(B\), \(B_s\), and \(B_c\) mesons, respectively. Notably, the \(c\)-quark and \(b\)-antiquark are close in the transverse plane for the \(B_c\)-meson. Additionally, as with charmed mesons, the distributions widen in \(x\) and narrow in \(b_\perp\) with increasing constituent quark mass, affecting the center of transverse momentum and maximum values.
\section{Summary}\label{sec:Summary}
Building on a recently proposed algebraic model that effectively describes the internal structure of the lowest-lying hidden-flavor pseudo-scalar and vector mesons (both light and heavy), we extended the approach to heavy-light pseudo-scalar mesons. The key advantage remains the same: the model allows for determining the leading-twist light-front wave function (LFWF) via its connection to the valence-quark parton distribution amplitude (PDA), which is often easier to compute using advanced continuum methods for solving QCD’s bound-state problems.

This algebraic model is based on constructing evidence-based ansätze for the meson’s Bethe-Salpeter amplitude (BSA) and quark propagator, enabling straightforward calculation of the Bethe-Salpeter wave function (BSWF). Projecting this onto the light front yields the LFWF, and integrating it over the transverse momentum squared leads to an algebraic link with the PDA. Using the known PDAs for the lowest-lying heavy-light pseudo-scalar mesons, we calculated their corresponding LFWFs, which were then used to derive generalized parton distributions (GPDs) in the DGLAP region. From these GPDs, we obtained related parton distribution functions (PDFs), electromagnetic form factors (EFFs), and impact-parameter space GPDs (IPS-GPDs). While experimental data and earlier theoretical predictions are limited, we have made comparisons where possible. We hope this study will inspire future experiments and theoretical advancements.

\end{document}